\documentclass[%
 reprint,
 amsmath,amssymb,
 aps,
]{revtex4-2}

\usepackage{graphicx}
\usepackage{dcolumn}
\usepackage{bm}
\usepackage{xcolor}

\begin{document}

\preprint{APS/123-QED}

\title{Synchronization of Bloch Oscillations in array of parallel Josephson Junctions }

\author{R. Shaikhaidarov}
  \email{r.shaikhaidarov@rhul.ac.uk}
	\affiliation{Royal Holloway, University of London, Egham Hill, Egham TW20 0EX, United Kingdom}
  
     \author{R. Hussain}
	\affiliation{Royal Holloway, University of London, Egham Hill, Egham TW20 0EX, United Kingdom}

    
    \author{V.N. Antonov}
	\affiliation{Skolkovo Institute of Science and Technology, Bolshoy Boulevard 30, Moscow, Russia}
 
    \author{E. Il’ichev} 
    \affiliation{Leibniz Institute of Photonic Technology, D-07702 Jena, Germany}

\date{\today}

\begin{abstract}
We demonstrate the synchronization of current quantization in a parallel array of weakly coupled Josephson Junctions operating in the regime of the coherent quantum phase slip. The first quantized current step on the voltage-current characteristic of 69 parallel Josephson junctions, under microwave excitation at a frequency of $f$ = 19.325~GHz, is observed at 426 nA. Experiments show that accuracy of quantization does not degrade with increasing number of combined Josephson junctions. This demonstration addresses the issue of quantized current amplitude requirements adopted for a practical quantum standard, while leaving the question of accuracy for further research. At present, the quantum current standard of accepted metrological accuracy in the triangle of electrical units is under development. The other two, the Volt and Ohm standards based on the Josephson and Quantum Hall effects, respectively, are already well established in metrology.          
\end{abstract}

\maketitle

In the early stages of the development of the quantum theory of solids, F. Bloch predicted the unusual dynamics of a quantum particle in a periodic potential under the action of a constant force \cite{Bloch1929}. Instead of the uniform motion expected by traditional classical physics, such a Bloch particle oscillates with a frequency proportional to the applied force. This prediction has now been well confirmed in experiments with a semiconductor superlattice \cite{Feldmann1992} and atoms in an optical potential \cite{Dahan1996}. It turned out that small Josephson junctions (JJ) obey a similar Bloch Hamiltonian, and one can expect the appearance of “Bloch oscillations” of the superconducting phase $\varphi$ \cite{Averin1985}. These oscillations can be synchronized by the externally applied microwave current $I_{\rm ac}$, leading to quantized current steps on voltage-current curves. These steps are developed at the $dc$ current 
\begin{equation}
 I_{dc,n} = n\times Q_0f,
 \label{eq:quantization}
\end{equation}
 
 \noindent where $Q_0=2e$ is the charge of Cooper pair, $f$ is the frequency of the applied microwave current, and $n$ is an integer. Recently, this quantum frequency-to-current converter  has been demonstrated in superconducting nanowires \cite{Shaikhaidarov2022} and JJs \cite{Shaikh2024,Kaap2024}. The effect is a direct consequence of the Coherent Quantum Phase Slip (CQPS): coherent flux tunneling across the superconducting weak link, the nanowire or JJ. CQPS is dual to the Josephson effect, a coherent tunneling of the Cooper pairs between two superconductors.

The first signature of current quantization in the JJ is the blockade of the current below the critical voltage $V_C=\pi E_S/e$ ($E_S$ is the CQPS energy) \cite{Shaikh2024}. When the microwave is applied to the JJ the current steps appear at the $I - V$ curves. The width of the $n$th step is determined by $V_C$ and microwave current amplitude $I_{\rm ac}$ as follows:
\begin{eqnarray}
\Delta V_n= 2J_n\left(\frac{I_{\rm ac}}{2e f} \right) V_C.
\label{eq:Bessel2}
\end{eqnarray}
\noindent Here $J_n$ are the Bessel functions of $n$th order. This result was obtained in \cite{Averin1985} for the limit $E_S \gg k_BT$. 
However, in experiments, the thermal energy of the environment, $k_BT$, is usually comparable to or exceeds $E_S$, $E_S \lesssim k_BT$. Then the critical voltage is normalized with the environmental noise as \cite{Shaikhaidarov2022}:      

\begin{eqnarray}
V_C^*= \frac{V_C^2}{8R\Delta I_{T}},
\label{volt}
\end{eqnarray}

\noindent where $R$ is the value of the normal resistance of the screening circuit and $\Delta I_{T}$ is the noise-induced current. The origin of $\Delta I_{T}$ is discussed later. 
We refer to $V_C^*$ as the apparent critical voltage. One can get $V_C^*$ from the blockade of the current in the $I-V$ curve measured with
the microwave switched off. With applied microwave, the environmental noise modifies  \eqref{eq:Bessel2} to \cite{Shaikhaidarov2022}: 

\begin{eqnarray}
\Delta V_n= 2J_n^2\left(\frac{I_{\rm ac}}{2e f} \right) V_C^*.
\label{eq:Bessel}
\end{eqnarray}

\noindent Thus, external noise reduces the width of the current plateaus due to the suppressed $V_C^*$ and square of the Bessel function. The typical amplitude of $V_C^*$  is few microvolts. The width of the quantized steps also decreases with  $n$, so that only steps with $n$ = 1, 2 and 3 are of practical interest. 

The maximum amplitude of the quantized current is limited by the apparent critical current $I_C^*$. The latter obeys $I_C^*\leq I_Z$, where the Landau-Zener tunneling current $I_Z$ marks the boundary of a high probability of transitions from the lower to higher Bloch energy bands \cite{ZAIKIN1988125, Kuzmin1996, Shaikh2024}: 
 
 \begin{eqnarray}
I_Z=\frac{\pi E_J}{16E_{CJ}}I_C.
\label{eq:Zener}
\end{eqnarray}

 \noindent In this equation $I_C = \pi\Delta/2eR_J$, $E_J$ and $E_{CJ}$ are the critical current, Josephson and Charging energies of the JJ. The typical value of $I_{C}$ when JJ is in the CQPS regime is $\sim$ 100 nA, and $I_C^*$ is below a quarter of this amplitude. In practice, the quantized current $I_{dc,n}$ in a single junction is below $\sim$ 10 nA (corresponding microwave frequency $f\sim$ 30 GHz). A further increase in quantized current is limited by $I_Z$. However, for practical applications in metrology, one needs a current of mA and above. We believe that this can be achieved with a parallel array of synchronized JJs. 
 
\begin{figure}
\includegraphics[width=8.5 cm]{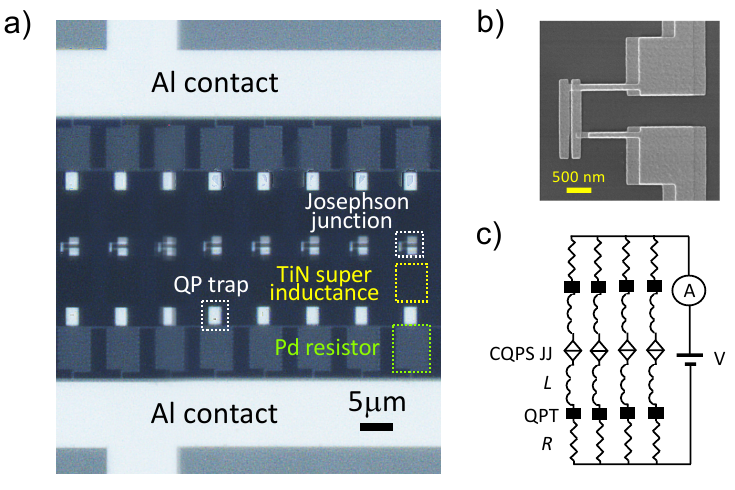}
\caption{\label{fig:fig1} a) Photo of a parallel array of JJs. Each JJ has screening circuit: TiN superinductance $L$, quasi-particle trap (QPT) and Pd resistors $R$; b) SEM image of 100$\times$100 nm$^2$ JJ; c) Electrical circuit diagram of the device.}
\end{figure}

Fifty years ago, a similar problem emerged when the voltage standard was developed. The voltage steps (Shapiro steps) $V$ = $n\times \Phi_0f$ appear in the $I-V$ curves of JJ when irradiated with the microwave of frequency $f$ ($\Phi_0 = h/2e$ is flux quantum). Although the maximum observe voltage of the Shapiro steps was in the millivolt range, experts from the US National Bureau of Standards proposed to use this effect for a new definition of volt \cite{Field1973}.  Due to relatively small generated voltages, the first single-junction Josephson standards were not practical. Research had focused on synchronization of the series array of JJs. It took time to improve junction reproducibility and optimize microwave - JJ coupling before the 10 V standard was released and implemented in several national standards laboratories \cite{Lloyd1987,Popel1991}.

In this Letter, we demonstrate a synchronization of quantized current in the array of parallel JJs operating in a regime of CQPS. The combined amplitude of the current plateaus in 69 JJs with $n$~=~1 is $\sim$ 426~nA. It exceeds recently reported 252~nA in the current standard based on  the Hall resistor (QAHR) and the programmable Josephson voltage (PJVS) standards \cite{Rodenbach2025}, while conceding to the latter in accuracy. We found that the width and accuracy of quantized plateaus do not degrade with the number of parallel JJ allowing a scaling of the parallel array of JJs. The limiting factors of accuracy are the thermal noise generated in the normal resistor of the screening circuit and the microwave heating.

Current quantization in the JJ is extremely delicate and sensitive to environmental noise. Each JJ in the array has its own protective (screening) circuit consisting of a compact 0.7~$\mu$H on-chip superinductor and a 23~k$\Omega$ Pd resistor connected symmetrically in series, Fig.~\ref{fig:fig1}. This technical implementation is similar to the one used for the protection of CQPS in individual JJs \cite{Shaikh2024,Kaap2024,Shaikhaidarov2022}. The impedance of such circuit varies from 23~k$\Omega$ at low frequencies to 67~k$\Omega$ at frequencies around 10 GHz. Further filtering and protection are taken by the $dc$ lines connecting the array to the measurement circuit. They are filtered by thermo-coax lines and cascade of LTCC low-pass filters with a stop band from 80 MHz to 20 GHz. The JJ array chip is housed in the copper box at the 15 mK stage of the dilution refrigerator.  The microwave signal is delivered through coaxial cables that are thermally encored at different temperature stages of the refrigerator. To suppress propagation of high frequency white noise the microwave line has series of attenuators distributed at different temperature stages of the refrigerator with total suppression of -40 dB.

\begin{figure}
\includegraphics[width=8.5 cm]{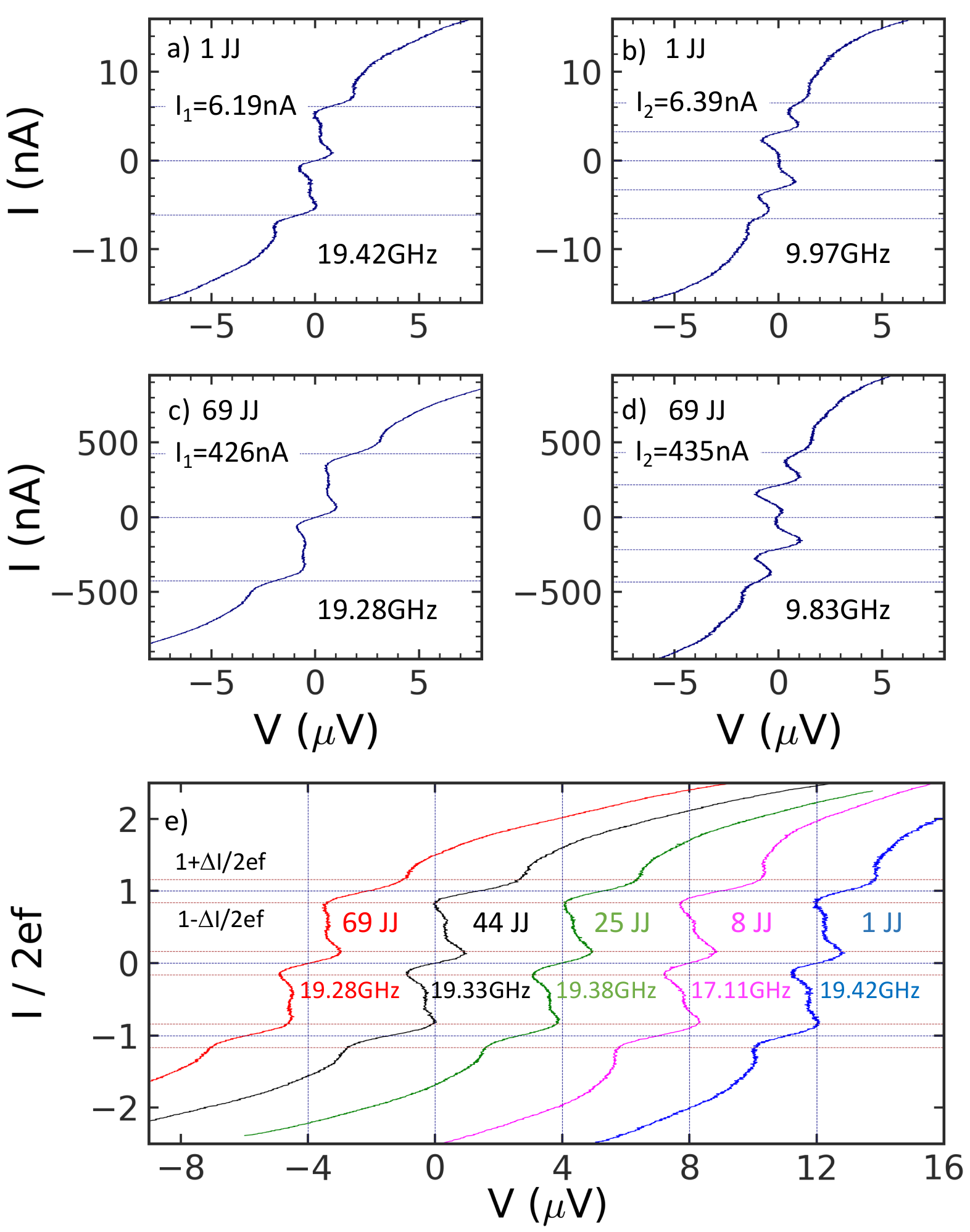}
\caption{\label{fig:fig2} Current quantization in a single JJ a) at 19.42~GHz, b) at 9.97~GHz  and in parallel array of 69 JJs c) at 19.28GHz, d) 9.83~GHz.  Horizontal guidelines correspond to multiple of  $I=N\times 2ef$ ($N$ = 1 in a), b) $N$ = 69 in c), d)); e) Quantized current steps in parallel arrays of JJs normalized by $2efN$. The $I-V$ curves are shifted horizontally by 4~$\mu V$ for clarity. There are extra horizontal guide lines $n\pm\Delta I/2ef$, where $\Delta I$=~1~nA is the deviation of current from the quantized value at the tip of the blockade voltage.}
\end{figure}

Samples are fabricated using standard nanotechnology processing, including electron beam lithography, reactive ion etching, and thin film deposition. There are four stages in the fabrication flow. We start with a Si wafer having 5 nm thick TiN film. In the first stage, Ti/Au coplanar waveguide for the microwave signal and the contact pads for the $dc$ signal are defined. In the second stage, a TiN superinductor is fabricated using reactive ion etching with resist mask. The third stage is the fabrication of compact Pd resistors. Finally, in the fourth stage, the Al/AlOx/Al Josephson Junctions are formed.

Arrays with parallel 10, 30, 50 and 80  junctions are studied. Each chip also has a single junction structure for assessing the scaling effect. The typical size of the JJs is 50 $\times$ 40 nm$^2$. They have tunneling resistance $R_J \sim$ 3~k$\Omega$ and capacitance $C_J \sim$ 10$^{16}$ F (capacitance of 100$\times$100 nm$^2$ aluminium JJ is $\sim$ 5$\times$10$^{-16}$ F). We estimate $E_C \sim$ 193~GHz and $E_J \sim$ 49~GHz in individual JJ in the array. 

The key experimental curves are shown in Fig.~\ref{fig:fig2}. Quantized current steps are present in the $I-V$ curve of a parallel array of nominally 80 JJs when exposed to microwaves, Fig.~\ref{fig:fig2} (c, d). The current quantization in a single junction is also shown as a reference in Fig.~\ref{fig:fig2} (a, b). When the microwave frequency is $f$ = 19.325~GHz, the first current step, $n=1$ in \eqref{eq:quantization}, $I_1 =2efN$=~426~nA corresponds to $N$ = 69 parallel junctions. The amplitude of the current in the array exceeds the current plateau in the individual JJ: it is significantly greater compared to $I_1$ =~6.19~nA in the single JJ. The same is true for the experimental data when the microwave frequency is changed to $f$ = 9.83~GHz. The second current plateau at this frequency is developed at $I_2=2\times 2efN$ =~435~nA.     

  \begin{figure}
\includegraphics[width=8 cm]{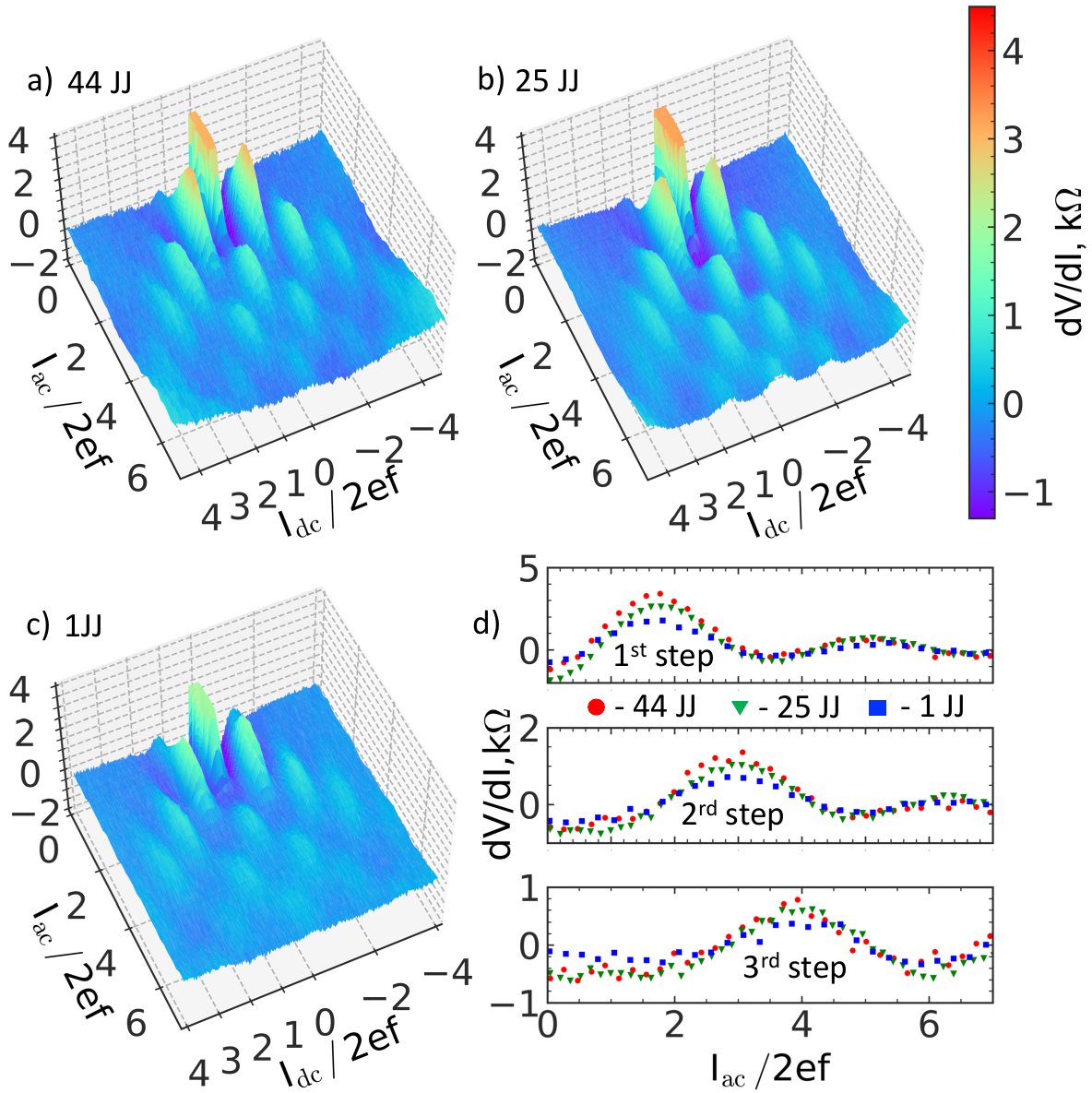}
\caption{\label{fig:fig3} The maps of the differential resistance $\text{d}V/\text{d}I$ in arrays of a) 44, b) 25 and c) single parallel junctions.  $dc$ and $ac$ currents are normalized by quantized current $2efN$ and $2ef$ respectively.  The $\text{d}V/\text{d}I$ is normalized by $N$; d) the slices of the maps a), b)  and c) along $I_{ac}$ axis at the position of quantized current with $n=1, 2 \text{ and } 3$. Data for the array with 44, 25, 1 JJs are colored with red, green and blue correspondingly.}
\end{figure}

The  $I-V$ curves with current normalized to the number of operating JJs in the arrays with different $N$ are shown in Fig.~\ref{fig:fig2} (e). The shape of the curves is identical to that of the single JJ. This indicates that the JJs are well synchronized and that the same microwave power is delivered to each JJ.

To demonstrate the Bessel-like behavior of the current steps in \eqref{eq:Bessel}, we record the differential resistance map $\text{d}V/\text{d}I$ in the coordinates of $I_\text{dc}$ and $I_\text{ac}$ for three arrays with $N$ = 44, 25 and 1, Fig.~\ref{fig:fig3}. $I_\text{dc}$ and $I_\text{ac}$ are normalized to the value of current quantization 2$ef$. $I_\text{dc}$ and differential resistance $\text{d}V/\text{d}I$ are also normalized to the number $N$ of JJs operating in the array. The maximum resistance at the centers of the first quantized plateau reaches $\sim 4$~k$\Omega$. The maps for single and arrays of JJs appear to be identical in shape. 
The cross sections of the map at the position of current quantization 2$efn$ with $n$ = 1,2 and 3 are shown in Fig.~\ref{fig:fig3}(d). The amplitude of differential resistance does not degrade with $N$. This gives further evidence of the feasibility of scaling the JJ array to obtain a sufficiently large quantized current.

Fairy good synchronization indicates close parameters of the JJ in the arrays. Consequently, the standard fabrication technology of aluminum JJ appears to be sufficient for this task. There are also requirements of the microwave frequency for the phase locking of JJ array. The question was addressed when the JJ voltage standard was developed \cite{Krautz1996}. In a series of JJ the microwave frequency should obey $f\gg f_p$, where  $f_p=\sqrt{2E_JE_{C}}/h$ is the plasma frequency of the individual JJ in the array. The dual requirement for the current standard takes the form $f\gg f_p=\sqrt{2E_SE_L}$, where $E_S$ is dual to $E_J$, and the inductive energy $E_L=\Phi_0^2 /2L$ is dual to $E_C$ \cite{Mooij06}. To calculate the inductive energy we use the dominating inductance of the screening circuit $L\simeq$~0.7~$\mu$H and obtain $E_L\simeq$ 4.6~GHz. The CQPS energy $E_S$ can be found using experimental $V_C^*\simeq$ 8 $\mu$V:

 \begin{eqnarray}
E_S=\frac{e\sqrt{8R\Delta I_T V_C^*}}{\pi},
\label{eq:CQPS energy}
\end{eqnarray}

\begin{figure}
\includegraphics[width=8.5 cm]{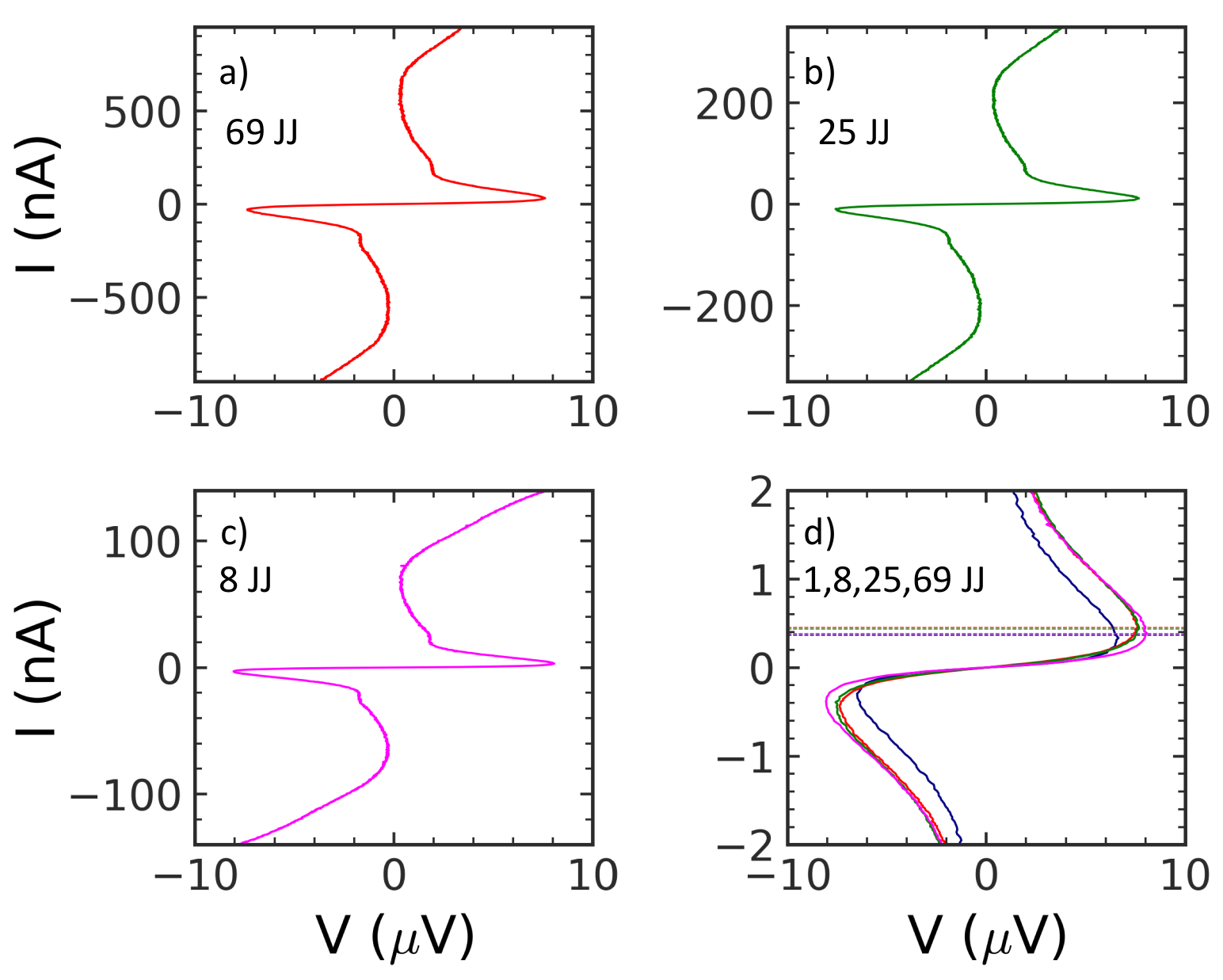}
\caption{\label{fig:fig4} $I-V$ curves in the arrays with different number $N$ of JJs  a) 69, b) 25, c) 8; d) $I-V$ curves normalized by $N$. The color scheme corresponds to a), b), c). The reference  $I-V$ curve of single JJ is blue. The thermal noise current is shown by the dashed horizontal line of corresponding color. Across all arrays it has value $\Delta I_T$~=~0.4$\pm$0.1~nA. The microwave is switched off.}
\end{figure}

 \noindent We estimate $E_S\simeq~$1.9~GHz when taking the resistance of the Pd meander and the thermal noise current $\Delta I_\text{T}\simeq$ 0.4~nA ($\Delta I_\text{T}$ is found experimentally, see Fig.~\ref{fig:fig4} and discussion). Then the plasma frequency is $f_p\simeq$ 4.2 GHz. It satisfies the phase locking criteria in the range of our experiment $f$~=~6~-~19~GHz. It is worth mentioning that $E_S$ is close to $E_L$ in our system, similar to the dual condition for the formation of optimal Bloch bands, $E_J\sim E_C$. 

The key question for the  application of the CQPS JJ array to metrology is the accuracy of quantization. In the ideal standard, the plateaus should be absolutely flat.  In our experiments a finite slope is clearly visible for all plateaus, regardless of the number of JJs in the array, see Fig.~\ref{fig:fig2}. To respond to a metrological requirements flatness of the quantized current plateaus should be better than one part per million. Such an accuracy presents a significant challenge. One can characterize the slope by the deviation from the quantized current at the tip of the blockade voltage, $\Delta I$. It is approximately 1~nA per JJ in Fig.~\ref{fig:fig2} e). The origin and evolution of $\Delta I$ can be obtained from analysis of the $I-V$ curves in the absence of microwave, see Fig.~\ref{fig:fig4}.

The curves have the current blockade below apparent critical voltage $V_C^*\simeq$~8~$\mu$V followed by a quasi-supercurrent region with apparent critical current $I_C^*\simeq$ 20~nA (for individual JJ). The latter is believed to be Zener current, see Eq.~\eqref{eq:Zener}.  Fig.~\ref{fig:fig4} presents data of arrays with 69, 25 and 8 parallel JJs, and combined graph with current normalized to $N$. The $I-V$ curve of a single JJ is included for reference. The normalized curves with different $N$ closely coincide in features, indicating that the CQPS parameters of the individual JJ in the array are close to each other, as we have already confirmed alternatively. The shape of the $I-V$ curves can be approximated with \cite{Shaikhaidarov2022} (Supplementary information):

\begin{equation}
 V(I_\text{dc})=\frac{V_C^2}{4R}\frac{I_\text{dc}}{I_\text{dc}^2+\Delta I_T^2},
 \label{eq:IVC1}
\end{equation}

\noindent where $\Delta I_{T}$ is the thermal noise current generated in the screening circuit . 

Experimentally $\Delta I_T$ can be found from the shape of the $I-V$ curve: it is a non-zero current at the tip of blockade voltage. Thermal noise currents in arrays with $N$~=~ 1, 8, 25, 69 normalized to $N$ are shown as horizontal lines of the corresponding colors in Fig.~\ref{fig:fig4}(d). Across all arrays, it is $\Delta I_{T}$~=~0.4$\pm$~0.4~nA. One can estimate $\Delta I_{T}$ from the equation:

\begin{equation}
 \Delta I_T=\frac{\pi k_BT}{2eR}.
 \label{eq:dI_T}
\end{equation}

\noindent For the base temperature of our experiment $T=20$~mK and $R$~=~23~k$\Omega$ we estimate $\Delta I_{T}\simeq$~0.2~nA, which is close to the experimental value. The thermal current can also be calculated from the width of the differential resistance at zero bias \cite{Shaikhaidarov2022}. We believe that the deviation of the current $\Delta I\sim$~1~nA from the quantized values in Fig.~\ref{fig:fig2}(e) has the same origin. An additional source of heat appears in the experiments with quantization as a result of dissipation of microwave energy in the normal resistor. From \eqref{eq:dI_T} we estimate the resistor temperature under microwave radiation $T\simeq$~160~mK. The effect of the thermal noise current can be suppressed by increasing $V_C$ by the factor exp$(-2eV_C/k_BT$). However, in this way, limitations are imposed by the technology of JJ fabrication \cite{Antonov2026}.  Alternatively, thermal noise can also be reduced by optimizing circuit heat dissipation \cite{Lucas2023} and delivering microwave radiation to the JJ in a more efficient way.

In conclusion, we demonstrate current quantization in arrays of parallel Josephson Junction protected by a screening circuit. In the array with 69 JJs a record quantized current of 426 nA is achieved. The current scales linearly with the number of JJ which enables to scale up arrays to get higher value. Importantly, the accuracy of current quantization does not degrade with scaling up and remains at the level of accuracy of the individual JJ.  However, the accuracy in the current design of the circuit is well below the metrological requirements for the standard of current. The thermal fluctuations in the resistors of the screening circuits are the main source of noise. They produce a noise current $\Delta I_T$, which gives rise to a finite slope of the quantized plateaus. The thermal noise current is linearly scaling up with number of channels as well. Thus, a key objective in realizing a practical quantum current standard is the development of a smart design for an individual JJ with a screening circuit with a focus on the reduced effect of thermal fluctuations.

\section*{Authors' contributions}
R.S., E.I. and V.N.A. conceived the experiments.
R.S., designed devices, fabricated samples, and conducted measurements.  R.H. contributed to fabrication and measurements.  R.S., E.I. conducted the simulation and analysis of the data and wrote the manuscript. All authors contributed to editing of the manuscript.

\begin{acknowledgments}
R.S. acknowledges support by Engineering and Physical Sciences Research Council (EPSRC) Grants No. EP/Y022637/1, UKRI4030 . E.I would like to thank the German Federal Ministry of Research, Technology and Space (BMFTR) for partial support under Grant No. 13N17121/NbNanoQ
\end{acknowledgments}

\bibliography{Scaling.bib}

\end{document}